\documentclass{aa}  

\usepackage{graphicx}
\usepackage{txfonts}
\usepackage{natbib}
\usepackage{longtable}
\usepackage{multirow}
\usepackage{color}
\usepackage{url}
\usepackage{mathtools}
\usepackage{amsmath}
\usepackage{soul}
\usepackage{cleveref}
\usepackage{adjustbox}
\sethlcolor{yellow}

\usepackage[pdfpagemode={UseOutlines},bookmarks=true,bookmarksopen=true,
   bookmarksopenlevel=0,bookmarksnumbered=true,hypertexnames=false,
   colorlinks,linkcolor={blue},citecolor={blue},urlcolor={red},
   pdfstartview={FitV},unicode,breaklinks=true]{hyperref}

\usepackage[switch]{lineno}
\usepackage{rotating}

\def\fermi{{\it Fermi}-LAT~}

\def\ergs{erg~s$^{-1}$~}

\def\phcms{photons~cm$^{-2}$~s$^{-1}$~}

\newcommand{\myemail}{angioni@mpifr-bonn.mpg.de}

\newcommand{\angstrom}{\mathrm{\mathring{A}}}

\newcommand{\sT}{\sigma_{\rm T}}
\newcommand{\p}{^\prime}

\newcommand{\g}{\gamma}
\newcommand{\gp}{\gamma^{\prime}}

\newcommand{\dD}{\delta_{\rm D}}

\newcommand{\psim}{\lower.5ex\hbox{$\; \buildrel \propto \over\sim \;$}}
\newcommand{\lbar}{\lower.0ex\hbox{$\; \buildrel
{\lower0.0ex \hbox{-}} \over\lambda  \;$}}

\newcommand{\PKS}{PKS\,0346$-$27}

\newcommand{\cm}{\mathrm{cm}}

\newcommand{\erg}{\mathrm{erg}}

\newcommand{\GeV}{\mathrm{GeV}}

\newcommand{\s}{\mathrm{s}}

\newcommand{\Kelvin}{\mathrm{K}}

\begin{document} 
\batchmode

   \title{The large gamma-ray flare of the Flat-Spectrum Radio Quasar PKS\,0346$-$27}

   \author{R.~Angioni
          \inst{1,2}
	  \and  
        R.~Nesci    
        \inst{3}
      \and
        J.\,D.~Finke
        \inst{4}
      \and
        S.~Buson
        \inst{2}
      \and
        S.~Ciprini
        \inst{5,6}
          }

   \institute{Max-Planck-Institut f\"ur Radioastronomie, Auf dem H\"ugel 69, 53121 Bonn, Germany, email: \myemail
   \and
   Institut f\"ur Theoretische Physik und Astrophysik, Universit\"at
   W\"urzburg, Emil-Fischer-Str. 31, 97074 W\"urzburg, Germany
  \and
   INAF/IAPS, via Fosso del Cavaliere 100, 00133, Roma, Italy
   \and
   U.S. Naval Research Laboratory, Code 7653, 4555 Overlook Ave. SW, Washington, DC 20375-5352, USA
   \and
   Space Science Data Center - Agenzia Spaziale Italiana, Via del Politecnico, snc, I-00133, Roma, Italy
   \and
   Istituto Nazionale di Fisica Nucleare, Sezione di Perugia, I-06123 Perugia, Italy
   }

   \date{Received 13 March 2019 / Accepted 19 June 2019}

 
  \abstract
   {}
   {In this paper, we characterize the first $\g$-ray flaring episode of the FSRQ PKS\,0346$-$27 ($z=0.991$), as revealed by \textit{Fermi}-LAT monitoring data, and the concurrent multi-wavelength variability observed from radio through X-rays.}
   {We study the long and short term flux and spectral variability from PKS\,0346$-$27 by producing $\g$-ray light curves with different time binning. We complement the \textit{Fermi}-LAT data with multi-wavelength observations from ALMA (radio mm-band), REM (NIR) and \textit{Swift} (optical-UV, X-rays). This quasi-simultaneous multi-wavelength coverage allowed us to construct time-resolved spectral energy distributions (SEDs) of PKS\,0346$-$27, and compare the broadband spectral properties of the source between different activity states using a one-zone leptonic emission model.}
   {PKS\,0346$-$27 entered an elevated $\gamma$-ray activity state starting from the beginning of 2018. The high-state continued throughout the year, displaying the highest fluxes in May 2018. We find evidence of short-time scale variability down to $\sim1.5$ hours, which constrains the $\g$-ray emission region to be compact. The extended flaring period was characterized by a persistently harder spectrum with respect to the quiescent state, indicating changes in the broadband spectral properties of the source. This was confirmed by the multi-wavelength observations, which show a shift in the position of the two SED peaks by $\sim2$ orders of magnitude in energy and peak flux value. As a result, during the high state the non-thermal jet emission completely outshines the thermal contribution from the dust torus and accretion disk. The broadband SED of PKS\,0346$-$27 transitions from a typical Low-Synchrotron-Peaked (LSP) to the Intermediate-Synchrotron-Peaked (ISP) class, a behavior previously observed in other flaring $\gamma$-ray sources. Our one-zone leptonic emission model of the high-state SEDs constrains the $\g$-ray emission region to have a lower magnetic field, larger radius, and higher maximum electron Lorentz factors with respect to the quiescent SED. Finally, we note that the bright and hard $\g$-ray spectrum observed during the peak of flaring activity in May 2018 implies that PKS\,0346$-$27 could be a promising target for future ground-based Cherenkov observatories such as the Cherenkov Telescope Array (CTA). The CTA could detect such a flare in the low-energy tail of its energy range during a high state such as the one observed in May 2018.}
   {}


   \maketitle
%

\section{Introduction}
Radio-loud Active Galactic Nuclei (AGN) are the most common astrophysical source class in the $\gamma$-ray sky. Their relativistic jets produced by the central black hole emit bright non-thermal radiation across a wide range of wavelengths and energies, from radio to TeV. The vast majority of $\gamma$-ray detected AGN are blazars, i.e. sources where the relativistic jet is aligned at a small angle with the observer's line of sight, leading to strong relativistic Doppler boosting and beaming effects. Blazars are historically divided into two subclasses: Flat-Spectrum Radio Quasars (FSRQs), showing strong broad emission lines in their optical spectra, and BL Lacertae objects (BL Lacs), which typically have weak or absent optical spectral features. There is also evidence that flat-spectrum radio-loud Narrow-Line Seyfert 1 (NLS1) galaxies host low-power blazar-type jets aligned with our line of sight~\citep{Berton2018}.
The broadband Spectral Energy Distributions (SED) of blazars show a typical double-peaked structure. The low-energy peak is produced by synchrotron emission from the relativistic electrons in the jet. The emission processes giving rise to the high-energy emission peak are somewhat less clear, with the simplest models invoking Inverse Compton (IC) emission from the same electrons producing the low-energy synchrotron emission (the so called \textit{leptonic} emission models), and more advanced models involving relativistic protons as well \citep[\textit{lepto-hadronic} emission models, see, e.g., ][]{2013ApJ...768...54B}. \citet{2010ApJ...716...30A} introduced an alternative classification scheme for blazars based on the peak frequency of the synchrotron component $\nu_\mathrm{peak}^\mathrm{syn}$. Sources with $\nu_\mathrm{peak}^\mathrm{syn}<10^{14}$ Hz are called Low-Synchrotron-Peaked (LSP), those with $10^{14}$ Hz\,$<\nu_\mathrm{peak}^\mathrm{syn}<10^{15}$ Hz are called Intermediate-Synchrotron-Peaked, and those with $\nu_\mathrm{peak}^\mathrm{syn}>10^{15}$ Hz are called High-Synchrotron-Peaked (HSP). FSRQs (and blazar-like NLS1s) are typically LSP, while the peak synchrotron frequency in BL Lacs can range across the three classes \citep{2010ApJ...716...30A}. While this is true when considering archival multi-wavelength data, some sources have shown extreme changes in their broadband spectral properties during flaring states. This implies that their place in classification schemes such as the ones mentioned above should be considered as time-dependent. Exemplary cases of time-dependent blazar classification are PKS\,2155$-$304~\citep{Foschini2008},  PKS\,1510$-$089~\citep{2011A&A...529A.145D}, PMN\,J2345$-$1555~\citep{Ghisellini2013}, 4C\,+49.22~\citep{2014MNRAS.445.4316C} and PKS\,1441+25~\citep{ahnen15}. 

PKS\,0346$-$27 (also known as BZQ~J0348$-$2749, \citealt{2009A&A...495..691M}), with coordinates R.A. = 57.1589354$^\circ$, Decl. = $-$27.8204344$^\circ$ \citep[J2000,][]{Beasley2002} is a blazar source located at a redshift $z = 0.991$ \citep{White1988}. It was first identified as a radio source in the Parkes catalog \citep{1964AuJPh..17..340B}, and \cite{White1988} classified it as a quasar based on its optical spectrum. It was later revealed as an X-ray source by \textit{ROSAT} \citep[][and references therein]{1999A&A...349..389V}. In the $\g$-ray band, it was not detected by the Energetic Gamma Ray Experiment Telescope \citep[EGRET,][]{tho93}, but it was detected by \textit{Fermi}-LAT and included in the \emph{Fermi}-LAT First Source Catalog~ \citep[1FGL,][]{1fgl}. In the latest catalog, the \emph{Fermi}-LAT Fourth Source Catalog~\citep[4FGL,][]{4FGL} it is associated with the $\gamma$-ray source 4FGL~J0348.5$-$2749. No elevated state of $\g$-ray activity from this source has been reported and studied in detail so far.

A near infrared (NIR) flare from PKS\,0346$-$27 was first reported based on data taken on 2017 Nov 14 \citep[MJD 58071, ][]{Carrasco2017}. A few months later, strong $\gamma$-ray flaring activity was reported on 2018 Feb 02 (MJD 58151) based on \fermi data \citep{Angioni2018}. The source was found in an elevated state, reaching a daily $\gamma$-ray ($>100$~MeV) flux more than 100 times larger than the average flux reported in the 3FGL, and a significantly harder spectrum with respect to the one reported in the 3FGL. This prompted multi-wavelength follow-up observations which revealed enhanced activity in the optical-NIR \citep{Nesci2018,Vallely2018}, ultraviolet (UV) and X-ray \citep{Nesci2018b}. The high-energy flare continued over the following months, showing a second, brighter peak on 2018 May 13 \citep[MJD 58251, ][]{Ciprini2018}, with a daily flux $\sim150$ times the 3FGL catalog level.

In this paper, we characterize the flaring activity of PKS\,0346$-$27, probing the temporal and spectral behavior in $\gamma$-rays. We investigate the changes in the broadband SED using quasi-simultaneous, multi-wavelength and archival data, and comparing the observations to theoretical models. In Section 2 we describe the data acquisition and analysis in the different bands. In Section 3 we discuss the variability observed with each instrument, with a focus on the object's $\gamma$-ray properties. In Section 4 we qualitatively describe the time-resolved quasi-simultaneous SEDs of PKS\,0346$-$27, and in Section 5 we discuss a theoretical modeling of the SEDs.

Throughout the paper we assume a cosmology with $H_0 =  73$ km s$^{-1}$ Mpc$^{-1}$, $\Omega_{\mathrm{m}}$ =   0.27, $\Omega_{\mathrm{\Lambda}}$ =   0.73~\citep{2011ApJS..192...18K}.

\section{Data set and analysis}

\subsection{\textit{Fermi}-LAT}
\label{sec:lat_data}
The Large Area Telescope (LAT) is a pair-conversion telescope, launched on 2008 June 11 as one of the two scientific instruments on board the \textit{Fermi Gamma-ray
Space Telescope} \citep{Atwood2009}. Its main energy range is
0.1-100 GeV, but its sensitivity extends down to 20 MeV and up to $>1$
TeV \citep{Ajello2017}. In the third LAT point source catalog, the 3FGL~\citep{Acero2015}, PKS~0346$-$27 is associated to the $\g$-ray source 3FGL~J0348.6$-$2748, with a 0.1-300 GeV flux of $(9\pm2)\times10^{-9}$ photons cm$^{-2}$ s$^{-1}$ and a power-law spectrum with index $2.4\pm0.1$.

We have used the Python package \texttt{Fermipy} \citep{Wood2017}
to analyze the \textit{Fermi}-LAT data. We use Pass8 event data~\citep{2013arXiv1303.3514A} and select photons of the \texttt{SOURCE} class, in a region of interest (ROI) of 10$^\circ$ $\times$ 10$^\circ$ square, centered at the position of the target source. We include in the ROI model all point sources listed in the 3FGL and located within 15$^{\circ}$ from the
ROI center, along with the Galactic \citep{2016ApJS..223...26A} and
isotropic diffuse emission (\href{https://fermi.gsfc.nasa.gov/ssc/data/analysis/software/aux/gll\_iem\_v06.fits}{\texttt{gll\_iem\_v06.fits}} and
\href{https://fermi.gsfc.nasa.gov/ssc/data/analysis/software/aux/is\_P8R2\_SOURCE\_V6\_v06.txt}{\texttt{iso\_P8R2\_SOURCE\_V6\_v06.txt}}, respectively). We perform a binned analysis with 10 bins per decade in
energy and 0.1$^{\circ}$ binning in space, in the energy range 0.1-300 GeV, adopting the instrument response functions \texttt{P8R2\_SOURCE\_V6}. A correction for energy dispersion was included for all sources in the model except for the Galactic and isotropic diffuse components.

We first perform a maximum likelihood analysis over the full time range considered here, i.e. 2008 Aug 04 15:43:36.000 UTC to 2018 Sep 01 00:00:00.000 UTC (MET~\footnote{Mission Elapsed Time, i.e. seconds since 2001.0 UTC.} 239557417 to 557452805). In the initial fit, we model the ROI sources adopting the spectral shapes and parameters reported in the 3FGL catalog. We fit as free parameters the normalization and spectral index of the target source, and the normalization of all sources within 5$^{\circ}$ of the ROI center, in addition to the isotropic diffuse and Galactic components. Since our data set more than doubles the integration time with respect to the 3FGL catalog, we look for potentially new sources, present in the ROI, with an iterative procedure. We produce a Test Statistic (TS) map. The TS is defined as $2\log(L/L_0)$ where
$L$ is the likelihood of the model with a point source at a given position, and $L_0$ is the likelihood without the source. A value of TS=25
corresponds to a significance of
4.2$\sigma$, when accounting for two degrees of freedom~\citep{Mattox1996}. A TS map is produced by including a test source at each map pixel and evaluating its significance over the current model. We look for significant (TS$>$25) peaks in the TS map, with a minimum separation of 0.3$^{\circ}$, and add a new point source to the model at the position of the most significant peak found, assuming a power-law spectrum. We then fit again the ROI, and produce a new TS map. This process is iterated until no more significant excesses are found~\footnote{The source finding process resulted in the addition of four point sources to the 3FGL starting model. None of these are close or bright enough to the ROI center to significantly affect the results on the target source.}. We also perform a best-fit position localization for the target source and all new sources with TS$>$25 found in the ROI.

In order to investigate the temporal variability of \PKS, we computed several light curves with different time binning (see Section~\ref{sec:lat_var}). In each light curve, we perform a full likelihood fit in each time bin. The number of free parameters in the  fit depends on the statistics in each bin. We first attempt a fit leaving the normalisation and index of the target source and the normalization of all sources in the inner $3^\circ$ of the ROI free to vary. If the fit does not converge or results in a non-detection (TS$<$25), we progressively restrict the number of free parameters in the fit in an iterative fashion. First we reduce the radius including sources with free normalization to 1$^\circ$, then we leave only the spectrum of the target source free to vary, and finally we fix all parameters to their average values except for the target source's normalization. We consider the target source to be detected if TS$>9$ in the corresponding bin, and the signal-to-noise ratio (i.e., flux over its error, or $F/\Delta F$) in that bin is larger than two. If this is not the case, we report a 95\% confidence upper limit.

\subsection{\textit{Swift}-XRT/UVOT}

PKS\,0346$-$27 was observed with the \textit{Neil Gehrels Swift Observatory} \citep{2004ApJ...611.1005G} as a Target of Opportunity (ToO), following the detection of the first $\g$-ray flare \citep{Angioni2018}. The \textit{Swift} data were analyzed using the tools available at the \href{http://www.asdc.asi.it/}{ASI Space Science Data Center (SSDC)}. The source has no nearby potential contaminants either in X-rays nor in optical, so it was possible to use the standard aperture photometry technique to evaluate the source parameters. Archival \textit{Swift}-XRT observations of this source are available on 2009 Mar 29 and 2009 May 18. Unfortunately, for both dates the exposure times were too low to perform a meaningful X-ray spectral fit. Spectral fits were possible during the high state in 2018, modeling the source emission with a single power law, the N$_H$ absorption frozen to the Galactic value 
\citep[$9.1\times10^{19}$ cm$^{-2}$,][]{2005A&A...440..775K}, and a number of energy bins ranging between 4 and 12. 

\textit{Swift}-UVOT images in six bands were obtained for most of the pointings. Aperture photometry for all bands was obtained using the on-line tool available at the \href{http://www.asdc.asi.it/}{ASI SSDC}, which provides magnitudes in the Vega system and fluxes in mJy. No reddening corrections were applied, which are of the order of a few hundredth magnitudes at most, and therefore comparable to the photometric errors. Given the source redshift ($z=0.991$), the interstellar UV absorption band of the host galaxy at $2200\,\angstrom$, if present, falls outside the UVOT range, and therefore does not affect these observations.

\subsection{REM}
The Rapid Eye Mount (REM) telescope is a 60 cm robotic instrument located at La Silla Observatory in Chile, and operated by the Istituto Nazionale di Astrofisica (INAF)~\footnote{URL: \href{http://www.rem.inaf.it/}{http://www.rem.inaf.it/}}. Follow-up observations of \PKS\ with REM started a few days after the first reported $\g$-ray flare in February 2018, using the ROSS2 camera for the $r$ and $i$ bands, and REMIR for the $J$ and $H$ bands. Time sampling was initially daily, then reduced to two days and weekly. Aperture photometry with IRAF/apphot was performed using nine nearby comparison stars from the 2MASS catalog with AAA quality flag, and twelve stars from the Fourth US Naval Observatory CCD Astrograph Catalog \citep[UCAC4, ][]{2013AJ....145...44Z}. The aperture diameter was set at twice the FWHM of the images, which was not the same in different nights. This is not critical because no nearby sources were present in any filter at our sensitivity level. Conversion from magnitudes to mJy for all REM bands was computed using the on-line tool from the Louisiana State University~\footnote{URL: \href{http://morpheus.phys.lsu.edu/~gclayton/magnitude.html}{http://morpheus.phys.lsu.edu/~gclayton/magnitude.html}}. No Galactic reddening corrections were applied.

\subsection{ALMA}

PKS\,0346$-$27 has been observed in the mm-band as a calibrator for the Atacama Large Millimeter/submillimeter Array (ALMA) since 2012. The flux density measurements in Bands 3 (84--116 GHz), 6 (211--275 GHz) and 7 (275--373 GHz) are publicly available as part of the ALMA calibrator catalogue~\footnote{URL: \href{https://almascience.eso.org/alma-data/calibrator-catalogue}{https://almascience.eso.org/alma-data/calibrator-catalogue}}.

\section{Multi-wavelength variability}

\subsection{\textit{Fermi}-LAT}
\label{sec:lat_var}
\begin{figure*}[!htbp]
\begin{center}
\includegraphics[width=\linewidth]{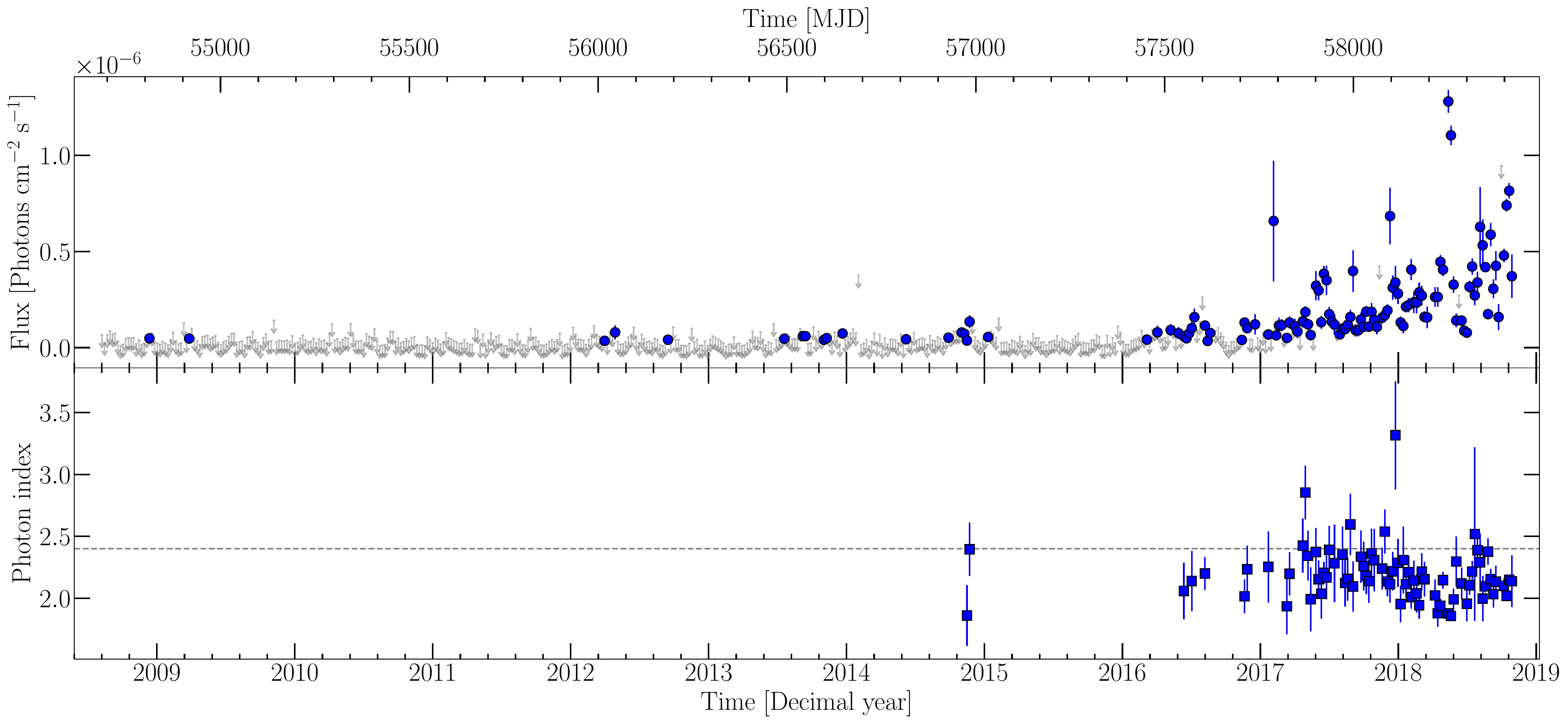}
\end{center}
\caption{Weekly binned \textit{Fermi}-LAT light curve of PKS\,0346$-$27 in the energy range 0.1-300 GeV, showing flux (top panel) and photon index (bottom panel). The latter is only plotted for bins where it was possible to fit it as a free parameter (see Section~\ref{sec:lat_data}). Filled blue points represent significant detections, downward grey arrows represent 95\% confidence level upper limits. The detection threshold is set at TS$>9$ and $F/ \Delta F>2$. For ease of representation, the length of the arrows has been set as equal to the average error for significantly detected bins. The dashed line in the bottom plot indicates the average photon index during the quiescent state (see Table~\ref{tab:lat}). A zoom-in of this light curve to the period after 2017.5 (MJD 57936.5) is shown in Fig.~\ref{fig:mwl_lc}.}
\label{fig:lc}
\end{figure*}

We show a weekly binned $\gamma$-ray light curve of PKS\,0346$-$27 in Fig.~\ref{fig:lc}. A pronounced peak is seen already in December 2017, lasting only one week. A more extended flaring period is seen between April and June 2018, with a peak 0.1-300 GeV flux of $(1.28\pm0.06)\times10^{-6}$~\phcms in the week centered on 2018 May 13. With a photon index of $1.88\pm0.04$, this translates to a luminosity of $(9.6\pm0.5)\times10^{48}$~\ergs. Finally, a period of renewed activity is observed again in October 2018, but not at the same magnitude as the events in May 2018.

\begin{figure*}[!htbp]
\begin{center}
\includegraphics[width=0.75\linewidth]{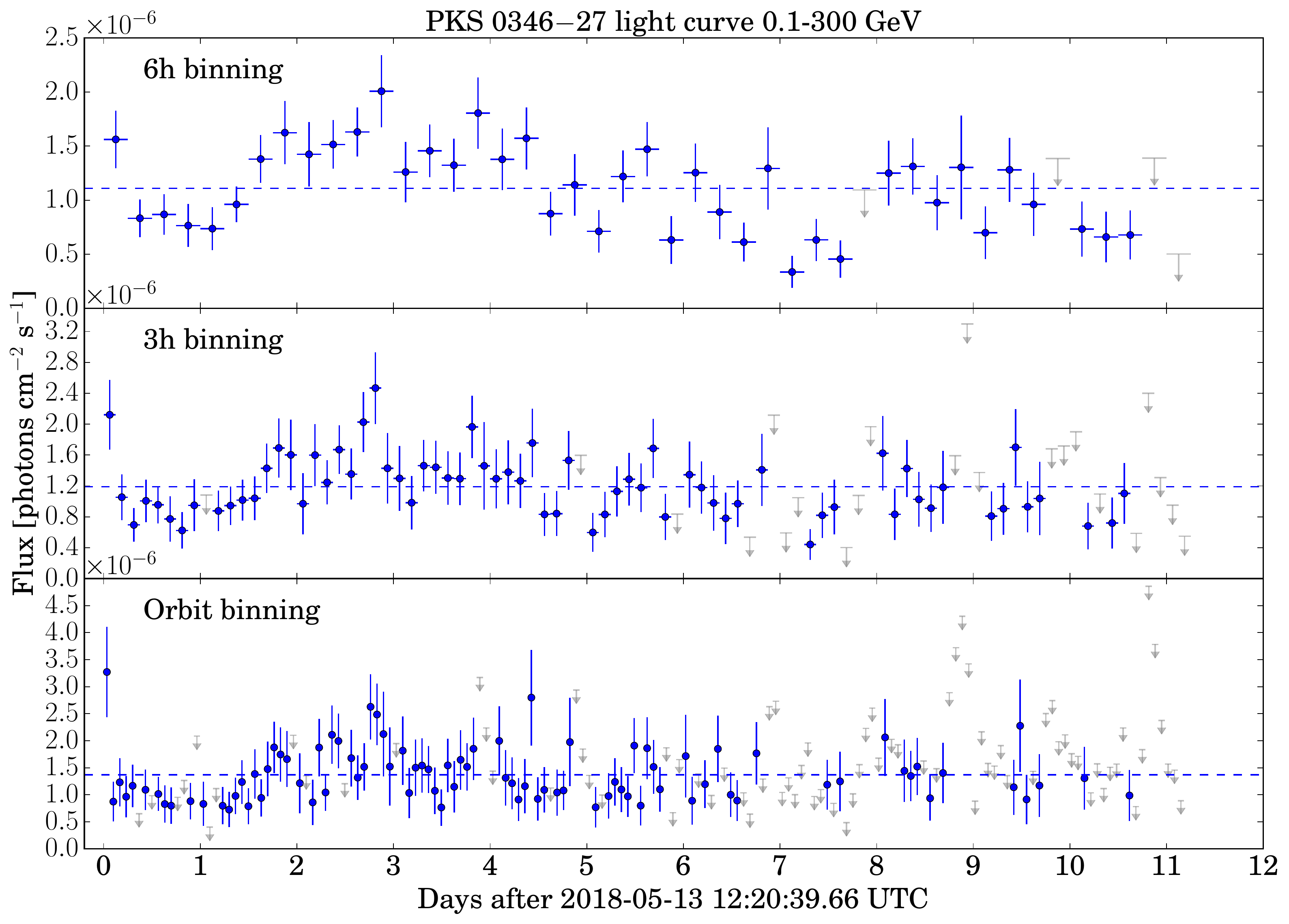}
\end{center}
\caption{Short-time scale \textit{Fermi}-LAT $\gamma$-ray light curves of PKS\,0346$-$27 in May 2018. Top to bottom panels: six hour, three hour, and orbital (95 minutes) binning, respectively. Filled blue points represent significant detections, downward grey arrows represent 95\% confidence level upper limits. The detection threshold is set at TS$>9$ and $F/ \Delta F>2$. For ease of representation, the length of the arrows has been set as equal to the average error for significantly detected bins in the corresponding light curve. The dashed blue lines represent the average flux considering only bins with significant detections.}
\label{fig:fast_lc}
\end{figure*}

We investigated the presence of significant short-time scale variability in PKS\,0346$-$27 during the brightest flaring period in May 2018. The 6-hour, 3-hour and orbit-binned light curves are shown in Fig.~\ref{fig:fast_lc}. The $\gamma$-ray flux of the source in this time period is high enough to yield significant detections in most bins, even on such short time scales. Fast variability appears evident at a visual inspection. To quantify its significance, we computed the so-called variability index~\citep[see][]{1fgl}. The variability index is defined as
\begin{equation}
    V = \sum_{i} w_i\,(F_i-F_{wt})^2
\end{equation}

where $F_{wt}$ is the weighted average and $w_i$ is the weight, which is computed as
\begin{equation}
    w_i = 1/\sigma_i^2
\end{equation}

for bins with a significant detection and 

\begin{equation}
    w_i = [(F^{UL}_{i}-F_i)/2]^{-2}
\end{equation}

for bins where an upper limit is placed ($F^{UL}$ is the 95\% confidence flux upper limit). The variability index follows a $\chi^2$ distribution with $N_\mathrm{bin}-1$ degrees of freedom. The results of the test are listed in Table~\ref{tab:chi2}.
\begin{table}[htbp]
\caption{Results of $\chi^2$ test for short-time scale variability for the light curves represented in Fig.~\ref{fig:fast_lc}.}
    \centering
    \begin{tabular}{cccc}
    \hline
    \hline
    Binning & $V/\mathrm{d.o.f.}$ & $p$-value & Significance\\
    \hline
    6 hours & 3.50 & $4\times10^{-14}$ & $11.7\,\sigma$\\
    3 hours & 2.66 & $3\times10^{-15}$ & $11.1\,\sigma$\\
    Orbital & 1.50 & $3\times10^{-5}$ & $4.6\,\sigma$\\
    \hline
    \hline

    \end{tabular}
    
    \label{tab:chi2}
\end{table}

The variability appears to be the most significant ($>10\sigma$) on 6 hours and 3 hours time scales, but is still strongly significant on orbital scales. In order to provide a more quantitative estimate of the shortest significant variability time scale, we have computed the doubling/halving time using two-point flux changes, following Eq.\,1 from \cite{Foschini2011}. In this procedure, we only consider bins with significant detections where the flux difference between consecutive bins (with no upper limit in between) is larger than two times their $1\sigma$ uncertainty (i.e., their error bar). In all three light curves, we find a shortest doubling/halving time scale consistent with the bin size.

\begin{figure}[!htbp]
\begin{center}
\includegraphics[width=\linewidth]{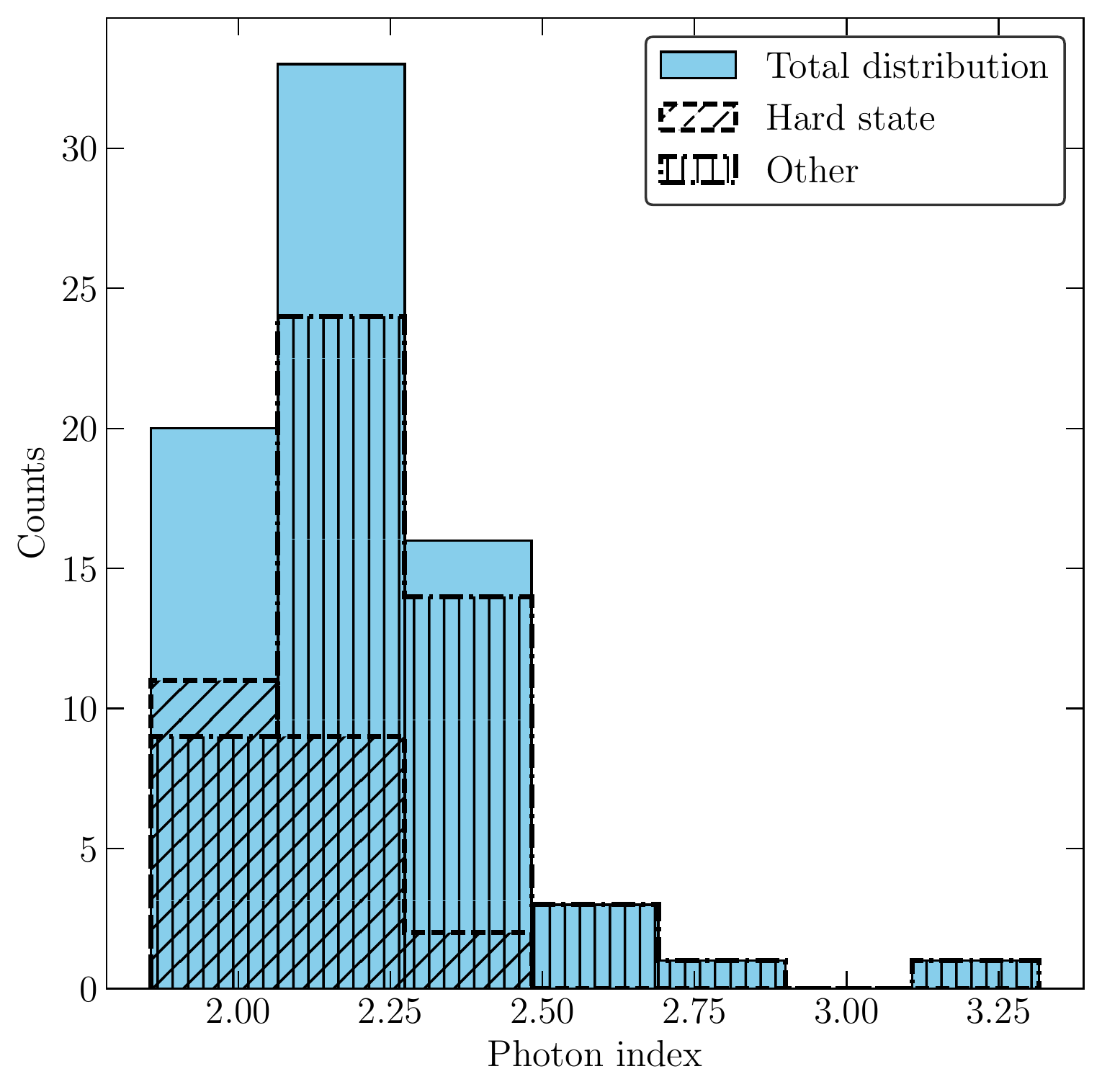}
\end{center}
\caption{Histogram of photon index from the \textit{Fermi}-LAT light curve shown in Fig.~\ref{fig:lc}, for bins when it was fitted as free parameter. The filled blue area represents the total distribution, while the hatched areas indicate the distributions for the ``hard state'' time interval and for all the other bins.}
\label{fig:index_hist}
\end{figure}

The variation in $\gamma$-ray photon index is shown in the bottom panel of Fig.~\ref{fig:lc}. It appears that the source transitioned to a different spectral state for the period approximately between January and July 2018 (2018.0-2018.55), showing a consistently harder spectrum with respect to the average value. We indicate this time interval as ``hard state''. There is a quite clear separation in the distribution of photon index from this time period compared to the rest of the light curve. This is best visualized with the histogram in Fig.~\ref{fig:index_hist}, showing the distribution of photon index. With respect to the average photon index excluding flaring periods (i.e., ``quiescent period'' in Section~\ref{sec:sed}), $\Gamma^\mathrm{index}_\mathrm{avg} = 2.43\pm0.07$, we can see that most of the points with $\Gamma^\mathrm{index}<\Gamma^\mathrm{index}_\mathrm{avg}$ are found during the hard state, while during the rest of the time the source spanned a much wider range of photon index, dominating the steeper end of the distribution. A two-sample Kolmogorov-Smirnov test yields a statistic of 0.38 and a $p$-value of $0.02$. Therefore the hypothesis that the distributions of photon index in the two time intervals are drawn from the same parent distribution can be rejected at $>95\%$ confidence level. 

\begin{figure*}[!htbp]
\begin{center}
\includegraphics[width=0.8\linewidth]{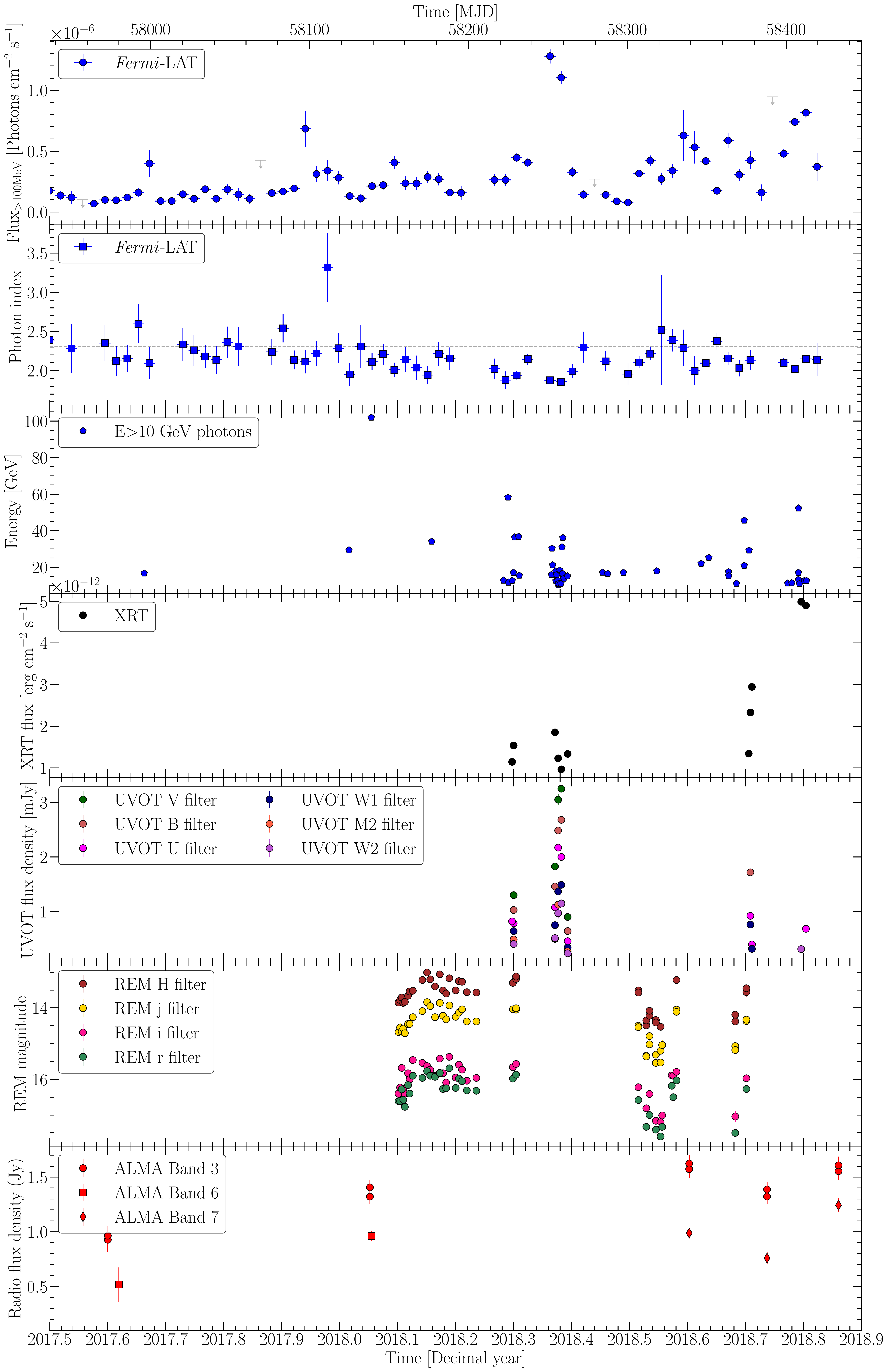}
\end{center}
\caption{Multi-wavelength light curves of PKS\,0346$-$27. Top to bottom: 0.1-300 GeV $\g$-ray flux (filled blue points represent significant detections, downward grey arrows represent 95\% confidence level upper limits; the detection threshold is set at TS$>9$ and $F/ \Delta F>2$), $\g$-ray photon index, $\g$-ray photons with energy larger than 10 GeV, \textit{Swift}-XRT flux between 2-10 keV, \textit{Swift}-UVOT optical-UV flux density, REM NIR magnitude, ALMA radio flux density.}
\label{fig:mwl_lc}
\end{figure*}

Finally, given the hard spectrum observed during the flare, we investigated the production of high-energy photons with $E>10$\,GeV. Using the \texttt{gtsrcprob} tool, we calculated the probability of each photon of being associated to PKS\,0346$-$27. In the third panel of Fig.~\ref{fig:mwl_lc}, we show the arrival time and energy of the highest energy photons with probability higher than 80\% of having been emitted by the target source. Interestingly, while most of the high-energy photons are clustered around the peak of activity in May 2018 (see Section~\ref{sec:sed}), the highest energy photon, with $E\sim100$~GeV, was observed on 2018 Jan 20.


\subsection{\textit{Swift}-XRT/UVOT}

The number of available \textit{Swift} observations is not sufficient for a detailed characterization of variability and higher-level processing such as cross-correlation analysis with the $\g$-ray data. The optical-UV and X-ray data all show enhanced emission with respect to the quiescent values. The XRT and UVOT light curves are shown in Fig.~\ref{fig:mwl_lc}. The resulting X-ray spectral parameters are listed in Table~\ref{tab:swift}. It is interesting to note that during the peak of the $\gamma$-ray activity in May 2018, PKS\,0346$-$27 was significantly more active in optical-UV than in X-rays. This reflects the shift in the position of the low-energy SED peak to higher frequencies in this period (see Sections~\ref{sec:sed} and~\ref{sec:mod}).

\begin{table*}[htbp]
\caption{\textit{Swift}-XRT/UVOT observations details.}
    \centering
    \begin{tabular}{lccccc}
    \hline
    \hline
    Date & Count rate (c/s) & Photon index & $\tilde{\chi}^2$(d.o.f.) & U$_\mathrm{mag}$ & V-W2 slope\\
    \hline
2009-03-29 & 0.017$\pm$0.003 & * & * & 17.66$\pm$0.06 & *\\
2009-05-18 & 0.012$\pm$0.003 & * & * & 18.06$\pm$0.10 & -0.54\\
2018-04-19 & 0.072$\pm$0.006 & 1.95$\pm$0.23 & 1.43(4)  &  15.66$\pm$0.02 & *\\
2018-04-20 & 0.073$\pm$0.004 & 1.72$\pm$0.11 & 1.11(10) &  15.70$\pm$0.02 &-0.13\\
2018-05-16 & 0.085$\pm$0.006 & 1.64$\pm$0.12 & 0.43(7)  &  15.36$\pm$0.01 & -0.37\\
2018-05-18 & 0.086$\pm$0.005 & 1.98$\pm$0.10 & 1.56(12) &  14.60$\pm$0.01 & -0.18\\
2018-05-20 & 0.081$\pm$0.005 & 2.34$\pm$0.14 & 1.04(8)  &  14.69$\pm$0.01 & -0.12\\
2018-05-24 & 0.042$\pm$0.004 & 1.53$\pm$0.27 & 2.33(2)  &  16.29$\pm$0.04 & -0.33\\
2018-09-15 & 0.103$\pm$0.008 & 2.01$\pm$0.20 & 0.60(4) & * & * \\
2018-09-16 & 0.097$\pm$0.008 & 1.56$\pm$0.17 & 1.32(4) & 15.53$\pm$0.02 & * \\
2018-09-17 & 0.061$\pm$0.006 & 1.09$\pm$0.25 & 0.41(2) & 16.44$\pm$0.03 & * \\
2018-10-18 & 0.121$\pm$0.008 & 1.51$\pm$0.13 & 0.84(7) & * & * \\
2018-10-21 & 0.147$\pm$0.009 & 1.70$\pm$0.13 & 0.69(9) & 15.85$\pm$0.01 & * \\

\hline
\hline

    \end{tabular}
    
    \label{tab:swift}
\end{table*}

\subsection{REM}
The source flux varied coherently in the four bands ($r$,$i$,$J$,$H$), showing significant variability also on a daily scale. These variations are fairly well correlated with the $\gamma$-ray ones. A mild correlation ($r=0.67$) between the ($r$-$H$) color index and the optical-NIR flux is present, hinting at a redder when fainter behavior, albeit with a large scatter. The average spectral slope in the range $r$-$H$ covered by REM is $-0.33$ in the plane log($\nu$)-log($\nu F \nu$).

\subsection{ALMA}
The $\gamma$-ray flaring activity in PKS\,0346$-$27 was accompanied by an enhanced state in the radio band as well, as revealed by the mm-band data from ALMA, shown in Fig.~\ref{fig:mwl_lc}. The flux density increased by a factor of three, rising from $\sim0.5$ Jy at the onset of activity up to $\sim1.5$ Jy at its peak. In this case as well, the sampling is not dense enough to perform a detailed analysis of correlated variability with other bands.

\section{Time-resolved SEDs}
\label{sec:sed}

\begin{figure*}[!htbp]
\begin{center}
\includegraphics[width=\textwidth]{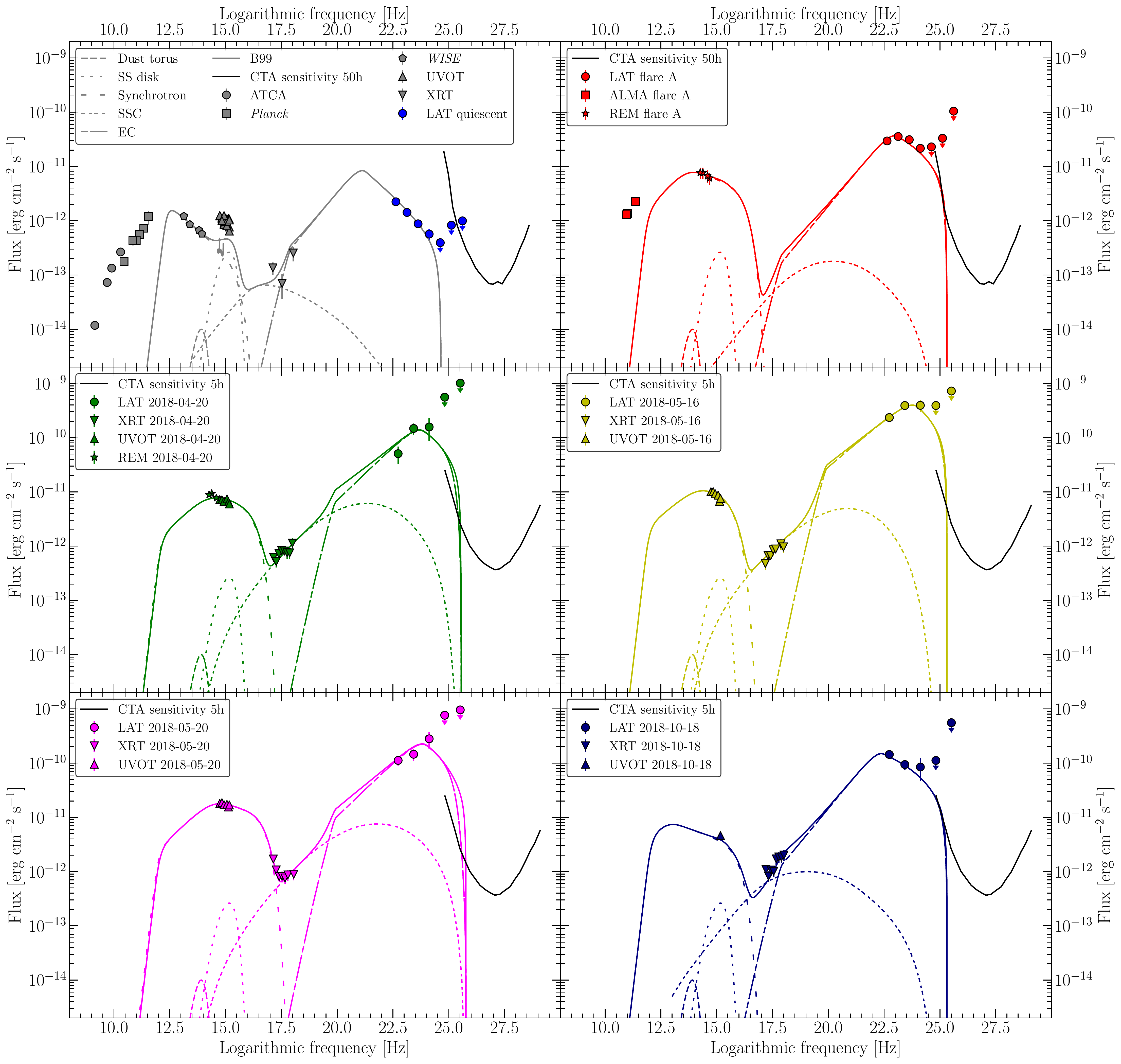}
\end{center}
\caption{Time-resolved SEDs of PKS\,0346$-$27. 
The color coding indicates different states of activity (see Section~\ref{sec:sed}). The corresponding activity states for each SED, indicated in the legend of each panel, are (top left to bottom right): quiescent, flare\,A, 2018 Apr 20, 2018 May 16, 2018 May 20 and 2018 Oct 18, respectively. The lines represent the resulting model SED for each state, where the solid line is the total emission, and the dashed line is the EBL-absorbed total emission. The different emission components included in the model are listed in the legend of the quiescent SED, and include the dust torus, Sakura-Sunyaev (SS) accretion disk, synchrotron, synchrotron self-Compton (SSC) and external Compton (EC) emission. In the quiescent SED, an adaptation of the optical spectrum from \cite{Baker1999} is included (B99 in the legend). The black solid lines indicate the CTA south sensitivity curves, for 50 hours of integration in the quiescent and Flare\,A states, and 5 hours of integration for the daily SEDs.}
\label{fig:sed_grid}
\end{figure*}

We have constructed time-resolved SEDs of PKS\,0346$-$27, by using contemporaneous data taken during the high-energy flaring state and archival data from different ground and space-based observatories. The time interval for each SED were chosen in order to probe different phases of activity, as traced by the $\gamma$-ray light curve. The multi-wavelength data for each time-resolved SED were selected based on the integration time used to produce the corresponding LAT spectrum. For example, if the LAT data for one SED were integrated over three months, all multi-wavelength data taken during this time interval are considered to be contemporaneous and are used to build the time-resolved SED. The SEDs are shown individually in Fig.~\ref{fig:sed_grid}, and the $\gamma$-ray properties in each of the states are summarized in Table~\ref{tab:lat}.
\begin{table}[htbp]
\caption{Results of \textit{Fermi}-LAT analysis on the different activity states of PKS\,0346$-$27.}
    \begin{center}
    \begin{tabular}{cccc}
    \hline
    \hline
    State & Flux$^\mathrm{a}$ & Photon index & TS\\
    \hline
    Quiescent & $1.3\pm0.1$ & $2.43\pm0.07$ & 289\\
    Flare A & $20\pm1$ & $2.07\pm0.04$ & 1278\\
    2018 Apr 20 & $50\pm10$ & $1.8\pm0.1$ & 180\\
    2018 May 16 & $170\pm10$ & $1.89\pm0.06$ & 1100\\
    2018 May 20 & $70\pm10$ & $1.8\pm0.1$ & 280\\
    2018 Oct 18 & $90\pm10$ & $2.2\pm0.1$ & 379\\
    \hline
    \hline
    \end{tabular}
    \end{center}
    $^\mathrm{a}$ Total flux in the energy range 0.1-300 GeV in units of $10^{-8}$ \phcms.
    \label{tab:lat}

\end{table}

The different activity states are defined as follows. For the quiescent state, we consider data until 2016 Jan 01, as this is the date after which the source started to be significantly detected on weekly time scales (see top panel of Fig.~\ref{fig:lc}). Since this was the first flaring episode observed in PKS\,0346$-$27, the quiescent \textit{Fermi}-LAT spectrum has been produced integrating from the mission start until 2016 Jan 01 (MJD 54682.66 to 57388). Together with the archival multi-wavelength data from the \href{http://www.asdc.asi.it/}{ASI SSDC}, this provides a well-sampled SED for the quiescent state of PKS\,0346$-$27, which is shown in the upper left panel of Fig.~\ref{fig:sed_grid}. The data show a well-defined synchrotron peak at a frequency of $\sim10^{12}$ Hz. A thermal emission component is clearly visible, peaking in the optical-UV region, which corresponds to the so called Big Blue Bump. The $\gamma$-ray spectrum is relatively steep, and combined with the X-ray data it constrains the high-energy SED peak to $\sim$MeV energies.

We compare this quiescent-state SED with several others constructed using multi-wavelength data taken after the start of the flaring activity in PKS\,0346$-$27. First, we defined a period of intermediate $\gamma$-ray activity between 2018 Jan 01 and 2018 Apr 01, which we indicate as ``flare A'' (MJD 58119 to 58209). During this time interval, the LAT spectrum is complemented by ALMA data in the mm-band and REM data in the NIR (see Fig.~\ref{fig:mwl_lc}). The corresponding SED is shown in the upper-right panel of Fig.~\ref{fig:sed_grid}. Given the large scatter in the REM data due to variability, we have averaged the data and taken its standard deviation as error, in order to be consistent with the averaged $\gamma$-ray spectrum. A shift in the SED peak positions is already evident in this intermediate state: the synchrotron peak lies now in the region $\sim10^{13}-10^{14}$ Hz. The combination of the peak shift and the change in normalization implies that the thermal component is now swamped by the non-thermal jet emission. The LAT spectrum is virtually flat, indicating a high-energy peak at GeV energies.

It is possible to construct an SED with data from the same day on 2018 Apr 20 (MJD 58228), probing a further higher activity period for PKS\,0346$-$27, intermediate between the long-term flaring and the flare peak, occurring in May. The daily SED is shown in the center-left panel of Fig.~\ref{fig:sed_grid}. The synchrotron peak is constrained by the REM and UVOT data, while the XRT and \textit{Fermi}-LAT data probe the high-energy peak. The synchrotron peak position is shifted further up to $\sim10^{14}$ Hz, and the simultaneous spectrum from NIR to UV is featureless, confirming that this region of the SED is completely dominated by non-thermal jet emission in this state. The most prominent change is the strong hardening of the $\gamma$-ray spectrum, suggesting a peak at $\sim10$\,GeV.

\textit{Swift} data allowed us to build an SED probing the week of highest flaring activity, on 2018 May 16 (MJD 58254), at a time when it was not possible to observe PKS\,0346$-$27 with ground-based facilities. The resulting daily SED is shown in the center-right panel of Fig.~\ref{fig:sed_grid}. The highest normalization of the high-energy peak is not accompanied by the hardest spectrum (see Table~\ref{tab:lat}), with a peak energy of the order of tens of GeV.

\begin{figure}[htbp]
    \centering
    \includegraphics[width = \linewidth]{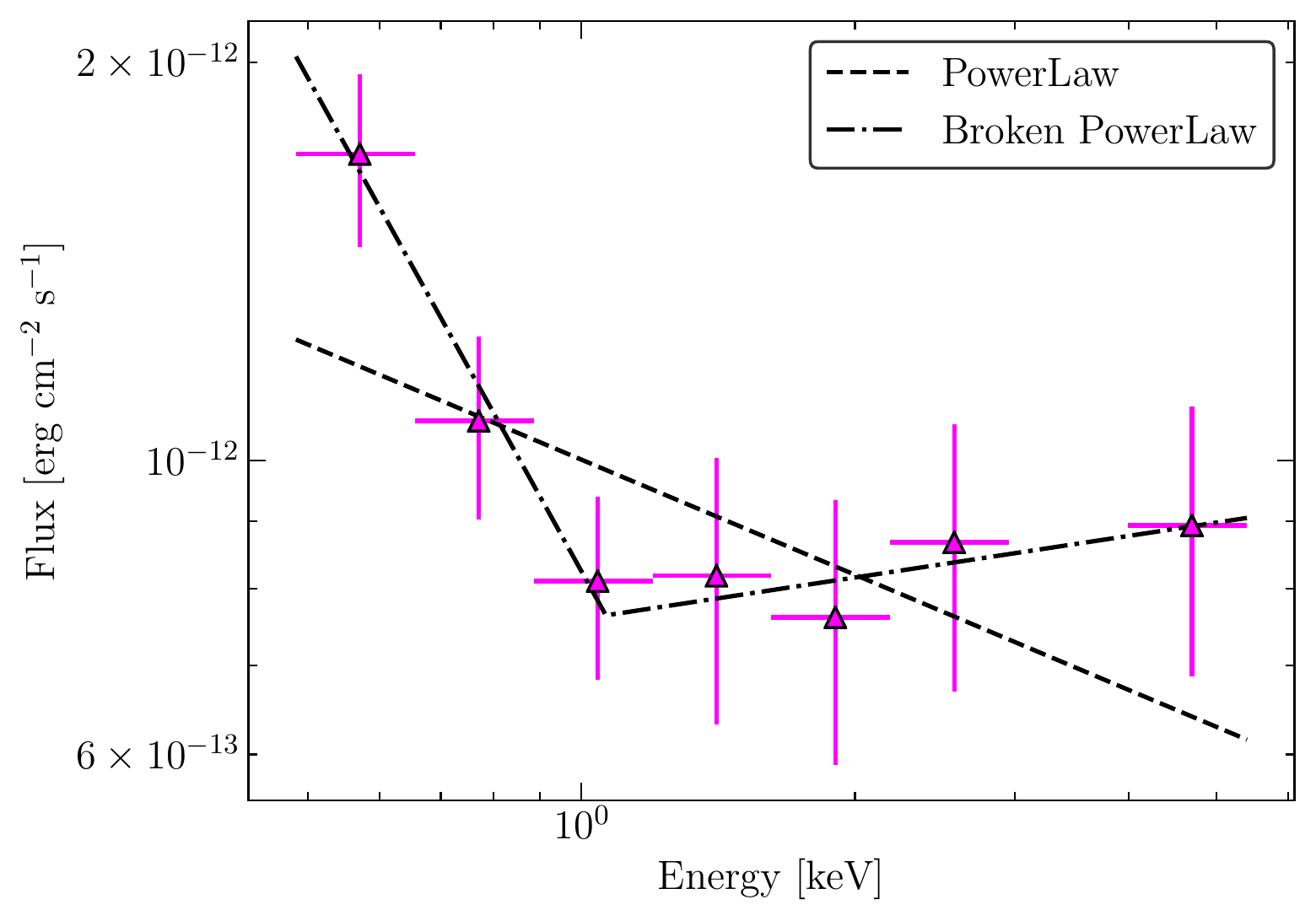}
    \caption{\textit{Swift}-XRT spectrum of PKS\,0346$-$27 on 20 May 2018. The dashed line indicates a power-law fit, while the dash-dotted line indicates a broken power-law fit.}
    \label{fig:xrt}
\end{figure}

Interestingly, another daily SED taken on the following week, on 2018 May 20, (MJD 58258, see lower-left panel of Fig.~\ref{fig:sed_grid}), shows a synchrotron peak position which appears to have shifted further right, to $\sim10^{14.5}$ Hz. Because of this, the X-ray data shows a clear synchrotron contribution at low energies, which causes a break in the spectrum. To better quantify the significance of this spectral feature, we have fitted the XRT spectrum from 2018 May 20 with a power-law (PL) and a broken power-law (BPL), using a simple $\chi^2$ regression. We defined the functions following their definition in XSPEC~\citep{1996ASPC..101...17A}, i.e.:

\begin{equation}\label{eq:pl}
    \frac{dN}{dE} = K\,E^{-\Gamma}
\end{equation}

\begin{equation}\label{eq:bpl}
\frac{dN}{dE}=\left\{
\begin{array}{ll}
K\,E^{-\Gamma_1} & \text{for } E\leq E_\mathrm{b}\\
K\,E_\mathrm{b}^{\Gamma_2-\Gamma_1}\,E^{-\Gamma_2} & \text{for } E>E_\mathrm{b}\\
\end{array} \right.
\end{equation}

\noindent for the PL and BPL models, respectively, where $K$ is the normalization at 1 keV, $\Gamma$ indicates the power-law photon index and $E_\mathrm{b}$ is the break energy. For the PL fit, we find $K=(6.3\pm0.5)\times10^{-4}$\,photons/keV/cm$^2$/s and $\Gamma = 2.3\pm0.2$, with a reduced chi-squared $\tilde{\chi}^2=1.687$ and five degrees of freedom. For the BPL fit, we find $K=(5.1\pm0.7)\times10^{-4}$\,photons/keV/cm$^2$/s, $\Gamma_1 = 3.2\pm0.4$, $\Gamma_2 = 1.9\pm0.3$, $E_\mathrm{b} = (1.1\pm0.2)$ keV, with a reduced chi-squared $\tilde{\chi}^2=0.126$ and three degrees of freedom~\footnote{The number of degrees of freedom is given by $\nu=N-k$ where $N=7$ is the number of bins and $k$ is the number of free parameters in the fit, i.e., $k=2$ and $k=4$ for the PL and BPL models, respectively.}. The XRT spectrum and the two models are shown in Fig.~\ref{fig:xrt}. An $F$-test to assess the significance of the BPL model with respect to the PL one yields $F=31.98$ and a $p$-value $p=0.0306$. The break is therefore significant at the 95\% confidence level.

Finally, we constructed a daily SED probing the period of renewed $\gamma$-ray activity in September-October 2018, namely on 2018 Oct 18 (MJD 58409, see lower-right panel of Fig.~\ref{fig:sed_grid}). While the normalization of the high-energy peak is comparable to the levels observed in May 2018, the spectrum is significantly softer (see Table~\ref{tab:lat}). Together with the hard X-ray spectrum, this indicates a peak in the $\sim100$ MeV range.


\section{SED modeling}
\label{sec:mod}
The quiescent and flaring states defined in Section~\ref{sec:sed} were modeled with the one-zone leptonic model described by \citet{finke08} and \citet{dermer09}. A brief description is given here.  The model assumes emission from an isotropic and homogeneous single spherical emitting region (``blob'') moving at speed $\beta c$ at small angle to the line of sight $\theta$ giving it a bulk Lorentz factor $\Gamma=(1-\beta^2)^{-1/2}$ and Doppler factor $\dD=[\Gamma(1-\beta\cos\theta)]^{-1}$.  We assume $\Gamma=\dD$. The blob has radius $R^\prime_b$ in the frame comoving with the blob and is filled with a tangled magnetic field of strength $B$. The model includes emission components from synchrotron, synchrotron self-Compton (SSC), and external Compton (EC) scattering of some external radiation field.  The external radiation field is assumed to be monochromatic, isotropic and homogeneous in the stationary frame, which is a reasonable approximation for scattering of a broad line region or dust torus \citep[e.g.,][]{dermer09,finke16}. We chose the external radiation field parameters to be similar to what one would expect for a dust torus.  Based on the dust model of \citet{nenkova08}, this implies the external radiation field has an energy density 

\begin{equation}
u_{\rm dust} = 2.4\times10^{-5}\ \left(\frac{\xi_{\rm dust}}{0.1}\right)
\left(\frac{T_{\rm dust}}{10^3\ \Kelvin}\right)^{5.2}\ \erg\ \cm^{-3}\ .
\end{equation}
Most blazars do not show significant signs of absorption in their $\gamma$-ray spectra from BLR photons~\citep[e.g., ][]{Costamante2018,Meyer2019}. PKS\,0346$-$27 does not show emission above 10 GeV, and thus it is not clear if these absorption features are present in its spectra. Nevertheless, we do not think models involving scattering of BLR photons are well-motivated any longer unless the SED of a blazar cannot be explained by other models. Emission components from a Shakura-Sunyaev accretion disk \citep{shakura73} and a blackbody dust torus were also included. 

The electron distribution in the emitting region was assumed to be a 
broken power-law
\begin{equation}
N_e(\gp) \propto \left\{ \begin{array}{ll}
\g^{\prime -p_1} & \gp_1 < \gp < \gp_{\rm brk} \\
\g^{\prime -p_2} & \gp_{\rm brk} < \gp < \gp_{2} \\
\end{array}
\right. 
\end{equation}
where $\gp$ is the electron Lorentz factor in the comoving frame.

The short-time scale light curves (Fig.~\ref{fig:fast_lc}), show significant variability down to $t_v \approx 6$\ hour (see Table~\ref{tab:chi2}). This constrains the size of the emitting region $R\p_b \la t_v \dD c/(1+z)$.  All our size scales in the modeling are consistent with this.

Our modeling results can be seen in Fig.~\ref{fig:sed_grid} and the model parameters
can be found in Table \ref{table_fit}. Based on the Mg\,II line measured from the optical spectrum of \cite{Baker1999} and the virial relations from \cite{Shen2011}, we estimate the black hole mass of PKS\,0346$-$27 to be $M_{\rm BH}\sim2\times10^{8}\ M_\odot$.

The small implied size scale $R\p_b$
indicates that significant synchrotron self-absorption occurs at radio
frequencies, so that the radio emission must be from a larger portion
of the jet \citep[e.g.,][]{blandford79,finke18}.  The absorption of
$\g$-rays by the extragalactic background light was included based on
the model of \citet{finke10}.

For the low state SED, we used a black hole mass consistent with the optical spectrum of \cite{Baker1999}. The disk ~\citep[modeled as a Shakura-Sunyaev disk, ][]{shakura73} and synchrotron emission were modeled so that their combined emission is in between the \cite{Baker1999} spectrum and the UVOT data, while reproducing the WISE data. Note that \cite{Baker1999} estimate a 20-50\% error on their spectrophotometric calibration. The UVOT data were likely taken when the source was in a higher state. The WISE and LAT quiescent state spectra give similar spectral indices, indicating that they are generated from the same electron population. The disk parameters are not well-constrained by the SED, but we use an inner disk radius of $6R_g$, which may not be consistent with a black hole with high spin.

For the Apr 20, May 16, and May 20 flares, the X-ray spectra cannot
connect smoothly with the LAT $\g$-ray spectra.  This indicates that
the X-rays are likely produced by SSC, and the soft excess in the May
20 X-ray spectrum indicates a contribution from the tail of the synchrotron component. The hard X-ray spectrum in the Oct 18 flare indicates it
is produced by the EC mechanism.  The lack of X-ray data during the
``flare A'' period implies that $R\p_b$ and other parameters are not well-constrained for this time period.

The constraints (EC from a dust torus, compact emitting region, X-rays
dominated by SSC during some flares) indicate that a large $\Gamma$ is
required to describe the SED, i.e. $\Gamma = 60$. Such a large value is consistent with the apparent speed of jet components seen with radio Very Long Baseline Interferometry (VLBI), which can be as high as $\beta_{\rm app}=78$~\citep{Jorstad2017}. In previous modeling efforts, it was
found that varying only the electron distribution was necessary to
reproduce several states from FSRQs
\citep[e.g.,][]{dammando13,ackermann14,buson14} although this is not
always the case \citep[e.g.,][]{dutka13,dutka17}.  For the SEDs of PKS\,0346$-$27, we attempted to model all states varying only the electron
distribution.  These efforts failed.  We then varied the size of the
emitting region and the magnetic field to model the states.  
Therefore, based on the classification of \citet{dutka13,dutka17}, all of the 
modeled flares from PKS\,0346$-$27 are type 2 flares~\footnote{\citet{dutka13,dutka17} classified flares into type 1, which can be modeled by changing the electron distribution only, and type 2, which require changes in other parameters such as the magnetic field strength or emission region size}.

The highest jet power is reached on 2018 May 16
and 2018 Oct 18, with $P_{\rm j,tot} = P_{\rm j,e} + P_{\rm j,B} =
5.0\times10^{45}\ \erg\ \s^{-1}$.  The accretion power is $P_{\rm acc}=L_{\rm disk}/\eta = 1.3\times10^{47}\ \erg\ \s^{-1}$ and the ratio $P_{\rm j,tot}/P_{\rm
  acc} = 0.04$.  The quiescent state and all flares except the ``flare A'' state have the jet power dominated by electrons;
however, as mentioned earlier, the model for the latter state is poorly constrained due to the lack of X-ray observations at this time.  The
general relativistic magnetohydrodynamic simulations by
\citet{tchekh11} indicate that the ratio $P_{\rm j,tot}/P_{\rm acc}$
can be as large as 1.4 if the jet is extracting power
from the spin of the black hole. The values of the total jet power calculated here can be considered lower limits, since we do not include the uncertain power carried by protons in the jet.

We computed the radiative cooling time scale in the observer frame at $1\ \GeV$ from the model parameters \citep{finke16},
\begin{equation}
t_{\rm cool} = \frac{ 3 m_ec^2 \sqrt{ 2(1+z)} }{4 c \sT [\Gamma^2u_{\rm seed} + B^2/(8\pi)]}
 \left( \frac{E_{\rm obs}}{1\ \GeV}\right)^{-1/2} 
\left(\frac{T_{\rm dust}}{2000\ \Kelvin}\right)^{1/2}\ ,
\end{equation}

\noindent and we list the results in Table \ref{table_fit}. The cooling time scale
should be less than the variability time scale, which is indeed the
case.

During all the flaring states, the peaks of both the synchrotron and
Compton components of the SED shift to higher frequencies.  This is
similar to the flares detected by the FSRQ PKS\,1441+25
\citep{abeysekara15,ahnen15}, although in that case the shift was even
more extreme.  For PKS\,1441+25 the synchrotron peak was as high as
$\approx 10^{15}$\ Hz, while for the flares from PKS\,0346$-$27
reported here the peak reaches a frequency of $10^{14.5}$\ Hz. The source PKS\,1441+25 was detected by MAGIC and VERITAS.  This fact, together with the the hard LAT spectra from PKS\,0346$-$27 during the Apr 20, May 16, and May 20 flares suggests that it might have been detectable by current imaging atmospheric Cherenkov telescopes such as H.E.S.S. (the only one able to access the relevant declination range). However, our modeling shows that the steeply falling high-energy peak would make a TeV detection challenging. To test whether PKS\,0346$-$27 would be a good target for the upcoming CTA~\citep{2017arXiv170907997C}, we plot the latest CTA sensitivity curves for CTA south~\footnote{From \href{https://www.cta-observatory.org/}{https://www.cta-observatory.org/}} in Fig.~\ref{fig:sed_grid}. It can be seen that a detection might be possible for states such as 2018 May 16 and 2018 May 20 in the low-energy tail of the CTA energy range, i.e. few tens of GeV. 


\begin{table*}
\footnotesize
\begin{center}
\caption{Resulting parameters from SED modeling.}
\label{table_fit}
\begin{tabular}{lccccccc}
\hline
\hline
Parameter & Symbol & Quiescent & Flare A & Apr 20 & May 16 & May 20 & Oct 18 \\
\hline
Redshift & 	$z$	& 0.991 & 0.991 & 0.991 & 0.991 & 0.991 & 0.991  \\
Bulk Lorentz Factor & $\Gamma$	& 60 & 60 & 60 & 60 & 60 & 60 \\
Doppler factor & $\delta_D$	& 60 & 60 & 60 & 60 & 60 & 60 \\
Magnetic Field [G]& $B$         & 3.1 & 2.6 & 1.2 & 0.82 & 1.4 & 1.3 \\
Comoving radius of blob [cm]& $R^{\prime}_b$ & $1.0\times$10$^{15}$ & $4.4\times$10$^{15}$ & $1.6\times$10$^{15}$ & $3.4\times$10$^{15}$ & $3.0\times$10$^{15}$ & $3.6\times$10$^{15}$ \\
\hline
Low-Energy Electron Spectral Index & $p_1$   & 2.0 & 2.0 & 2.2 & 2.3 & 2.2 & 2.0 \\
High-Energy Electron Spectral Index  & $p_2$ & 3.7 & 3.3 & 3.2 & 3.3 & 3.2 & 3.3  \\
Minimum Electron Lorentz Factor & $\gamma^{\prime}_{min}$  & $1$ & $1$ & $10$ & $10$ & $10$ & $1$ \\
Break Electron Lorentz Factor & $\gamma^{\prime}_{brk}$ & $50$ & $410$ & $1.1\times10^3$ & $1.3\times10^3$ & $1.3\times10^3$ & $20$ \\
Maximum Electron Lorentz Factor & $\gamma^{\prime}_{max}$  & $3.0\times10^3$ & $8.5\times10^3$ & $1.3\times10^4$ & $8.7\times10^3$ & $2.0\times10^4$ & $8.7\times10^3$ \\
\hline
Black hole Mass [$M_\odot]$ & $M_{BH}$ & $2\times10^8$ & $2\times10^8$ & $2\times10^8$ & $2\times10^8$ & $2\times10^8$ & $2\times10^8$ \\
Disk luminosity [$\erg\ \s^{-1}$] & $L_{disk}$ & $1.4\times10^{45}$ & $1.4\times10^{45}$ & $1.4\times10^{45}$ & $1.4\times10^{45}$ & $1.4\times10^{45}$ & $1.4\times10^{45}$ \\
Accretion efficiency & $\eta$ & $0.4$ & $0.4$ & $0.4$ & $0.4$ & $0.4$ & $0.4$ \\
Inner disk radius [$R_g$] & $R_{in}$ & $6.0$ & $6.0$ & $6.0$ & $6.0$ & $6.0$ & $6.0$ \\
Seed photon source energy density [$\erg\ \cm^{-3}$] & $u_{seed}$ & $3.7\times10^{-4}$ & $3.7\times10^{-4}$ & $3.7\times10^{-4}$ & $3.7\times10^{-4}$ & $3.7\times10^{-4}$ & $3.7\times10^{-4}$ \\
Dust Torus luminosity [$\erg\ \s^{-1}$] & $L_{dust}$ & $7.0\times10^{43}$ & $7.0\times10^{43}$ & $7.0\times10^{43}$ & $7.0\times10^{43}$ & $7.0\times10^{43}$ & $7.0\times10^{43}$ \\
Dust Torus radius [cm] & $R_{dust}$ & $6.9\times10^{17}$ & $6.9\times10^{17}$ & $6.9\times10^{17}$ & $6.9\times10^{17}$ & $6.9\times10^{17}$ & $6.9\times10^{17}$ \\
Dust temperature [K] & $T_{dust}$ & $2000$ & $2000$ & $2000$ & $2000$ & $2000$ & $2000$ \\
\hline
Jet Power in Magnetic Field [$\erg\ \s^{-1}$] & $P_{j,B}$ & $7.0\times10^{43}$ & $3.6\times10^{45}$& $1.0\times10^{44}$ & $2.1\times10^{44}$ & $4.7\times10^{44}$ & $6.0\times10^{44}$ \\
Jet Power in Electrons [$\erg\ \s^{-1}$] & $P_{j,e}$ & $2.5\times10^{45}$ & $5.1\times10^{44}$ & $2.9\times10^{45}$ & $4.8\times10^{45}$ & $2.4\times10^{45}$ & $4.4\times10^{45}$ \\
Observer Cooling time scale [s] & $t_{\rm cool}$ & $8.0\times10^2$ & $8.6\times10^2$ & $5.0\times10^2$ & $1.0\times10^3$ & $9.6\times10^2$ & $1.0\times10^2$ \\
\hline
\hline
\end{tabular}
\end{center}
\end{table*}

\section{Summary and conclusions}
In this paper we reported on the $\gamma$-ray flaring activity from the FSRQ PKS\,0346$-$27, and the associated multi-wavelength follow-up observations. The available data set, including mm-band radio, NIR, optical, UV, X-ray and $\gamma$-ray data, allowed us to investigate variations in the broadband SED of the source, and to probe its physical jet parameters through theoretical modeling of the SED in different activity states.

PKS\,0346$-$27 entered an elevated activity state between the end of 2017 and the beginning of 2018, with the flaring activity continuing over the whole year. We investigated the presence of fast variability during the brightest phase of flaring activity, and found significant variability down to time scales of hours, constraining the emission region to be very compact. Throughout 2018, the source showed a consistently harder spectrum with respect to the quiescent state, suggesting a drastic change in its broadband spectral properties. 

We investigated this by modeling the SED of the source in its quiescent state and comparing it to the one observed during several different stages of activity. The resulting time-resolved SEDs confirm that the multi-wavelength flare coincided with strong broadband spectral variability, with the two SED peaks shifting up by $\sim2$ orders of magnitude in energy during the peak of activity. While a clear thermal signature from the accretion disk is visible in the low-energy peak of the quiescent SED, during the flare this component is completely washed out by the non-thermal jet emission. According to our one-zone leptonic modeling, the high-state SEDs require a lower magnetic field, larger emission region size, and higher electron Lorentz factors. 

The time-resolved SEDs of PKS\,0346$-$27 show another example of a blazar with variable spectral classification. According to \cite{2010ApJ...710.1271A}, PKS\,0346$-$27 in its quiescent state would be classified as Low-Synchrotron-Peaked (LSP) blazar, as its low-energy SED peak is located at $\nu^\mathrm{syn}_\mathrm{peak}<10^{14}$ Hz. However, the shift in the synchrotron peak position during the flaring state puts it in the range of Intermediate-Synchrotron-Peaked (ISP) objects. Such a behavior has been observed before in the case of the LSP blazar PKS\,1441+25, which showed an HSP-like SED during the elevated state which led to its TeV detection~\citep{ahnen15}. Another example of a similar time-dependent SED classification is the FSRQ 4C\,+49.22 \citep{2014MNRAS.445.4316C}. Similarly to what we observe for PKS\,0346$-$27, 4C\,+49.22 showed a two orders of magnitude shift in its synchrotron peak position during a multi-wavelength flaring episode in 2011, transitioning from LSP to ISP. Moreover, the fact that the non-thermal emission is dominant in the optical band during the flare suggests that PKS\,0346$-$27 could possibly appear as a ``masquerading BL Lac'' in its high state, i.e., a blazar with intrinsically strong broad optical lines which are diluted by the non-thermal continuum, to an extent that would lead to its classification as a BL Lac based on its optical spectrum~\citep{2012MNRAS.420.2899G,2013MNRAS.431.1914G}.

The hard $\gamma$-ray spectrum suggests that PKS\,0346$-$27 could have been observed by Cherenkov telescopes in the TeV regime; however, the SED modeling shows a steep cutoff in the high-energy peak after $\sim10$ GeV. By comparing the model with current sensitivity curves from the upcoming CTA, we conclude that such a flare from PKS\,0346$-$27 might have been visible in the low-energy tail of the CTA energy range. Incidentally, if PKS\,0346$-$27 was detected in the TeV range, it would be the highest redshift VHE source ($z=0.991$), surpassing the gravitationally lensed blazar S3~0218+35 \citep[$z=0.944$,][]{2016A&A...595A..98A} as well as PKS\,1441+25 \citep[$z=0.940$,][]{ahnen15}. Because of this, detecting PKS\,0346$-$27 at VHE would also be relevant for EBL studies.

Continued monitoring of the GeV sky by the \textit{Fermi}-LAT is crucial in order to observe more flaring events from high-redshift ($z\gtrsim1$) blazars and establish the duty cycles of $\g$-ray activity in relativistic jets. Moreover, the \textit{Fermi}-LAT is an invaluable tool in order to trigger pointed observations by ground-based $\g$-ray observatories such as Cherenkov telescopes, including the upcoming CTA. Therefore, continued operations of the \textit{Fermi}-LAT into the CTA era would be instrumental in order to gain insight into the physics of blazar jets, both in the local universe and at cosmological distances.

\begin{acknowledgements}
We thank the anonymous journal referee, the \textit{Fermi}-LAT internal referee, Michael Kreter, the MPIfR internal referee, Laura Vega Garc\'ia, as well as Vaidehi Paliya, Philippe Bruel and David Thompson for useful comments which improved the manuscript.

This research made use of Astropy,\footnote{\href{http://www.astropy.org}{http://www.astropy.org}} a community-developed core Python package for Astronomy \citep{astropy:2013, astropy:2018}.

The \textit{Fermi}-LAT Collaboration acknowledges generous ongoing support from a
number of agencies and institutes that have supported both the
development and the operation of the LAT, as well as scientific data
analysis. These include the National Aeronautics and Space
Administration and the Department of Energy in the United States; the
Commissariat \'a l’Energie Atomique and the Centre National de la
Recherche Scientifique/Institut National de Physique Nucl\'eaire et de
Physique des Particules in France; the Agenzia Spaziale Italiana and
the Istituto Nazionale di Fisica Nucleare in Italy; the Ministry of
Education, Culture, Sports, Science and Technology (MEXT), High Energy
Accelerator Research Organization (KEK), and Japan Aerospace
Exploration Agency (JAXA) in Japan; and the K. A. Wallenberg
Foundation, the Swedish Research  Council,
and  the  Swedish  National  Space  Board in Sweden.
Additional support for science analysis during the operations
phase is gratefully acknowledged from the Istituto Nazionale di
Astrofisica in Italy and the Centre National d'Etudes Spatiales in
France. This work was performed in part under DOE Contract DE-AC02-76SF00515.

\end{acknowledgements}

\bibliographystyle{aa}
\bibliography{aa}

\end{document}